\documentclass[conference]{IEEEtran}
\usepackage{graphicx}
\usepackage[section]{placeins}
\usepackage{caption}
\usepackage{subcaption}
\usepackage{amsmath}
\usepackage{breqn}
\usepackage{lettrine}
\usepackage{stfloats}
\usepackage{psfrag}
\ifCLASSOPTIONcompsoc
  \usepackage[nocompress]{cite}
\else
  \usepackage{cite}
\fi

\ifCLASSINFOpdf
\else
\fi
\begin{document}
%
\title{A 2-6 GHz Ultra-Wideband CMOS Transceiver for Radar Applications}

\author{Alin Thomas Tharakan, Prince Philip, Gokulan T, Sumit Kumar, Gaurab Banerjee\\
\{alintharakan, princephilip, gokulant, sumitkumar13, banerjee\}iisc.ac.in\\
Department of Electrical Communication Engineering, Indian Institute of Science, Bangalore, India, 560012

}


\IEEEcompsocitemizethanks{\IEEEcompsocthanksitem M. Shell was with the Department
of Electrical and Computer Engineering, Georgia Institute of Technology, Atlanta,
GA, 30332.\protect\\
E-mail: see http://www.michaelshell.org/contact.html
\IEEEcompsocthanksitem J. Doe and J. Doe are with Anonymous University.}
\thanks{Manuscript received April 19, 2005; revised August 26, 2015.}

%
%

\markboth{Journal of \LaTeX\ Class Files,~Vol.~14, No.~8, August~2015}%
{Shell \MakeLowercase{\textit{et al.}}: Bare Advanced Demo of IEEEtran.cls for IEEE Computer Society Journals}
%



\IEEEtitleabstractindextext{%
\begin{abstract}
This paper presents a low power, low cost transceiver architecture to implement radar-on-a-chip. The transceiver comprises of a full ultra-wideband (UWB) transmitter and a full UWB band receiver. A design methodology to maximize the tuning range of the voltage-controlled oscillator (VCO) is presented. At the transmitter side, a sub-harmonic mixer is used for signal up-conversion. The receiver low noise amplifier (LNA) has a 2 to 6 GHz input matching bandwidth with a power gain of 9 dB and a noise figure of 2.5 dB. The transceiver is implemented in Cadence EDA tools using 65nm CMOS technology. The system achieves a total dc power consumption of 50 mW. Good noise figure performance; good wide-band matching; gain; high level of integration; low power; low cost of the proposed UWB radar transceiver front-end make it a highly competitive SoC solution for low power UWB transceivers.
\end{abstract}

}

\maketitle

\IEEEdisplaynontitleabstractindextext

%
\IEEEpeerreviewmaketitle

\ifCLASSOPTIONcompsoc
\IEEEraisesectionheading{\section{Introduction}\label{sec:introduction}}
\else
\section{Introduction}
\label{sec:introduction}
\fi

\lettrine[findent=3pt]{\textbf{T}}The rapid advancements in microelectronics, with technologies like 130 nm SiGe and 65 nm CMOS, have enabled the integration of entire millimeter-wave transceivers onto a single chip, promising low-cost solutions. However, these systems face challenges in transmit power, phase noise, and amplifier performance.

Since 2002, when the FCC allocated the 3.1–10.6 GHz band for commercial use, ultra-wideband (UWB) has transitioned into a major research area. UWB technology uses wide spectrum and short Gaussian pulses, offering advantages such as high data throughput, low power, resistance to interference, and simpler transceiver design. It excels in radar sensing and can penetrate objects better than narrowband signals, making it ideal for short-range, high-speed data transmission, indoor video, sensor networks, and positioning applications. These advantages make UWB an alternative solution to the narrow-band technology for sensing applications.

\begin{figure}
  \includegraphics[width=8cm,height=8cm]{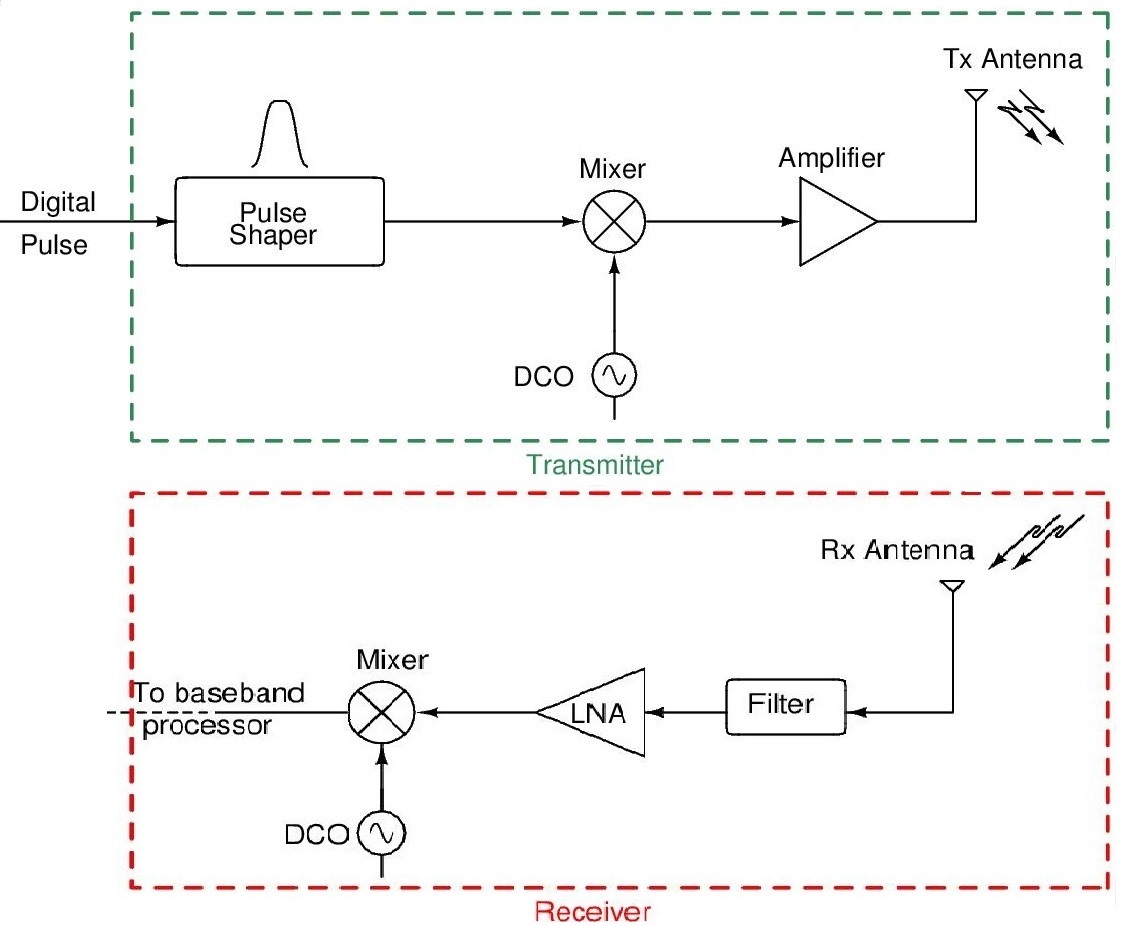}
  \caption{Transceiver Architecture}
  \label{trans_arch}
\end{figure}

UWB has gained attention for high-precision radar and target positioning due to its wide bandwidth. The transceiver architecture includes a digital pulse generator and a Pulse Shaping Circuit for Gaussian pulse generation, followed by an Up-conversion mixer for transmission. At the receiver, the signal is filtered, amplified with a Low Noise Amplifier, and down-converted for digital baseband processing.

The paper is organized as follows. In section II, the design of various transmitter blocks is considered. In section III, the design of the wideband LNA is summarized. Section IV discusses the results after implementation in Cadence EDA.

\section{Transmitter}
The UWB transmitter includes a Digitally Controlled Oscillator (DCO), a Pulse Shaping Circuit, and an Up-Conversion Mixer. A digital pulse is shaped and modulated to the desired frequency band. The design of these blocks is detailed below.

\subsection{Pulse Shaping Circuit}
\begin{figure*}[!t]
  \centering
  \subfloat[]{\includegraphics[width=0.442\linewidth]{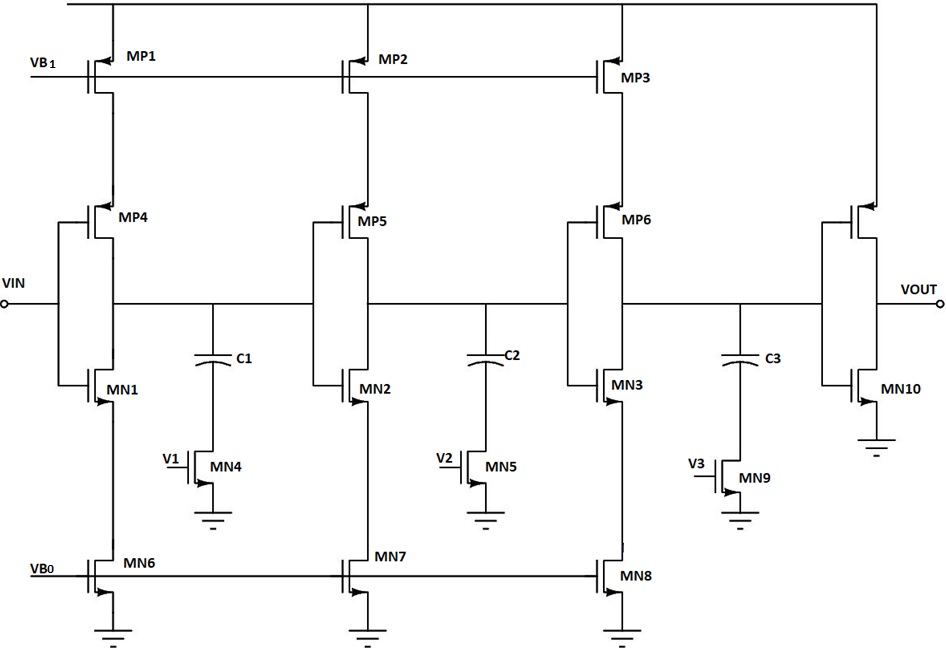}
		\label{pulse_shaper}}
 \subfloat[]{\includegraphics[width=0.29\linewidth]{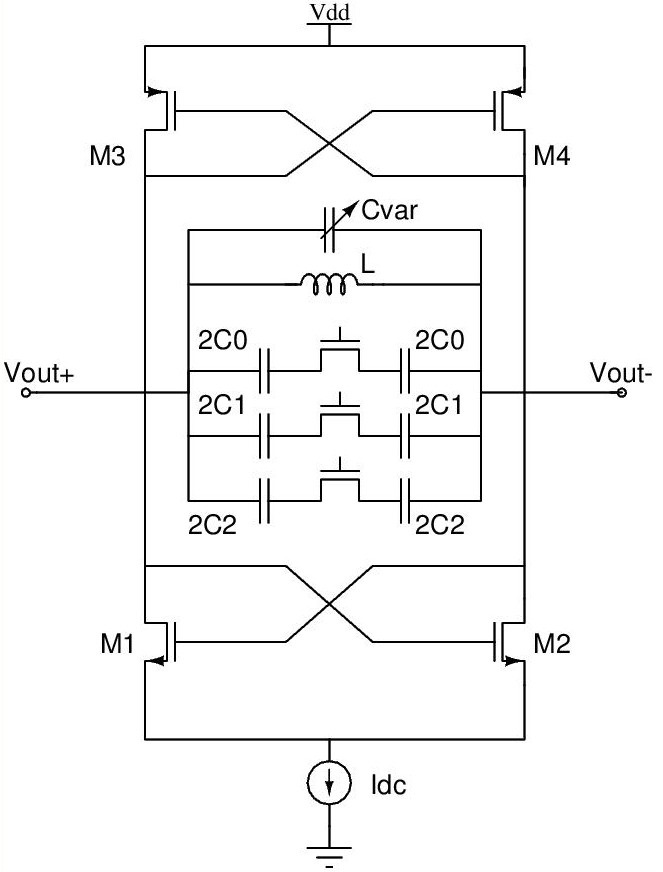}
		\label{dco}}  
  	\subfloat[]{\includegraphics[width=0.3\linewidth]{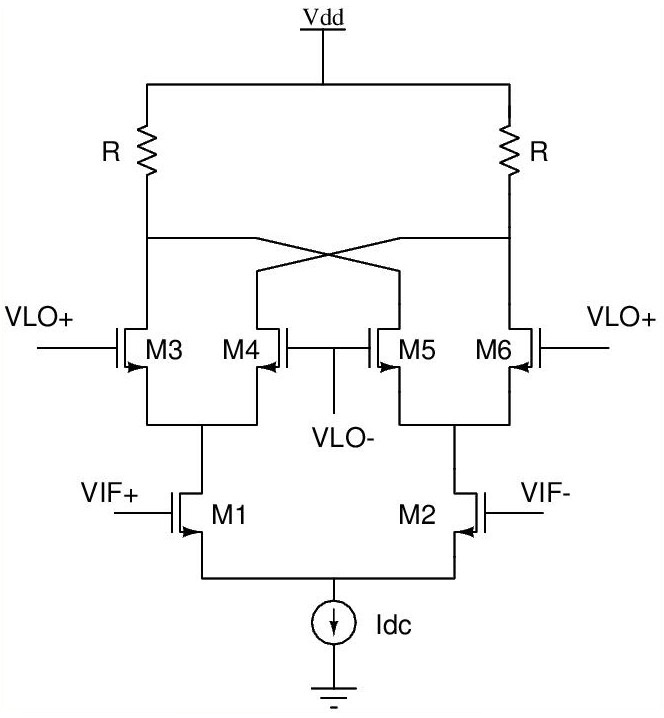}
		\label{mixer}}  
        
	\subfloat[]{\includegraphics[width=0.28\linewidth]{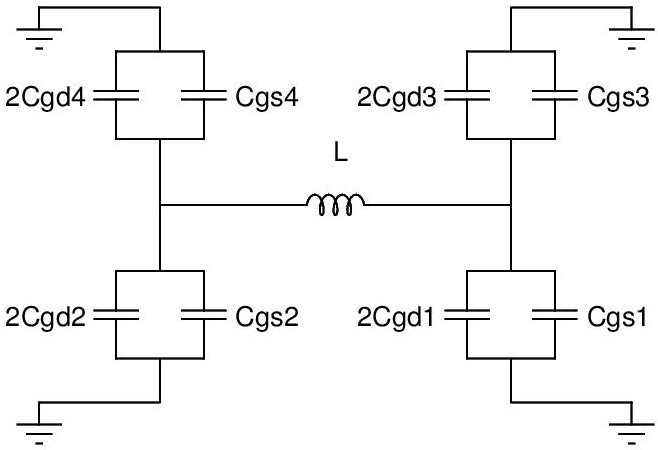}
		\label{vco_par}}        
	\subfloat[]{\includegraphics[width=0.35\linewidth]{{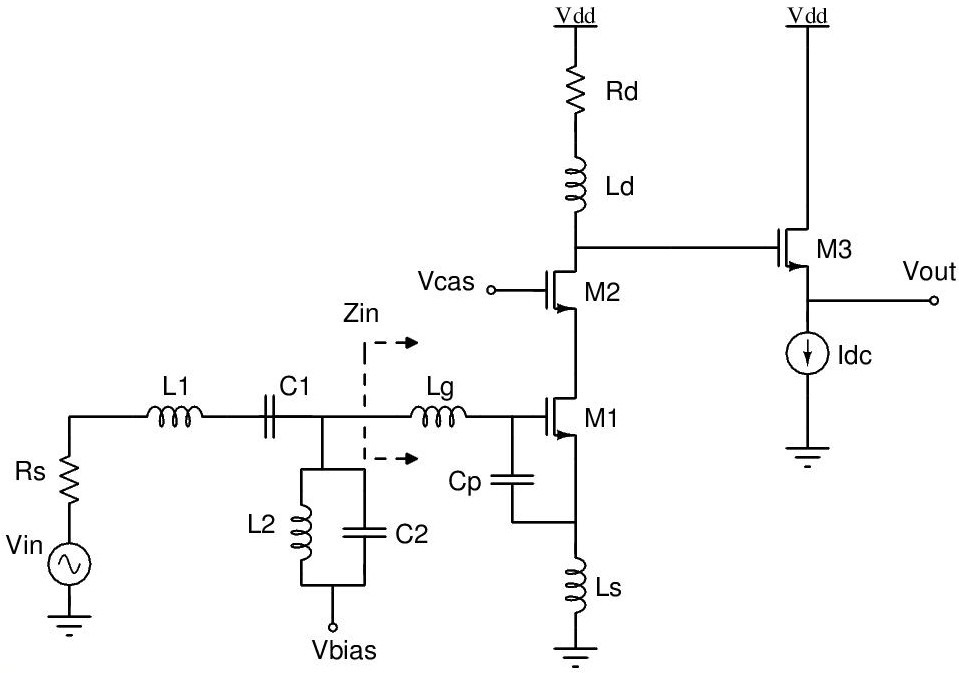}}
		\label{LNA_C}}
  	\subfloat[]{\includegraphics[width=0.35\linewidth]{{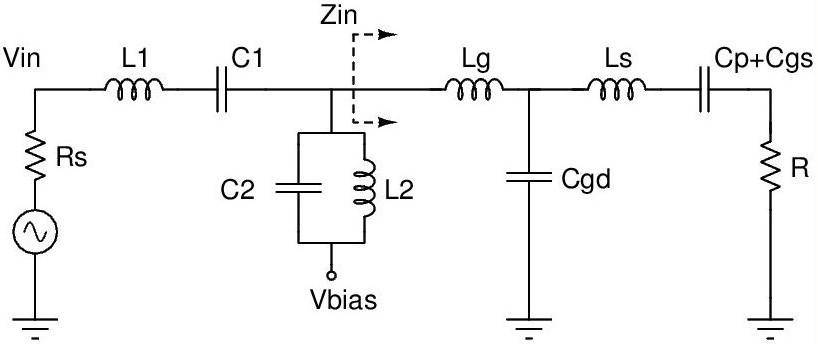}}
		\label{LNA_i}}
  \caption{\protect\subref{pulse_shaper}~Pulse shaping circuit. \protect\subref{dco}DCO circuit; 
 \protect\subref{mixer}~Up-conversion mixer; \protect\subref{vco_par}~Parasitic capacitance across L; \protect\subref{LNA_C}~LNA circuit diagram; \protect\subref{LNA_i}~LNA input network}
  \label{taucellf}
\end{figure*}

\IEEEPARstart For UWB radar, spectral efficiency and pulse localization are crucial. The Gaussian pulse excels in time-bandwidth product and meets stringent mask requirements better than a rectangular pulse, as it concentrates power within the main lobe. Fig.\ref{pulse_shaper} illustrates the circuit for generating an approximate Gaussian pulse\cite{pulse_shaper}. In the circuit, transistors MN6, MN7, MN8, MP1, MP2, and MP3 act as current sources. The bias voltages $V_{B0}$ and $V_{B1}$ control inverter current via the \textit{Current Starving technique}, while capacitors $C_1$, $C_2$, and $C_3$ enable pulse shaping.
\FloatBarrier

 \begin{figure*}[!t]%
    \centering
    \subfloat[\centering  ]{{\includegraphics[scale=0.32]{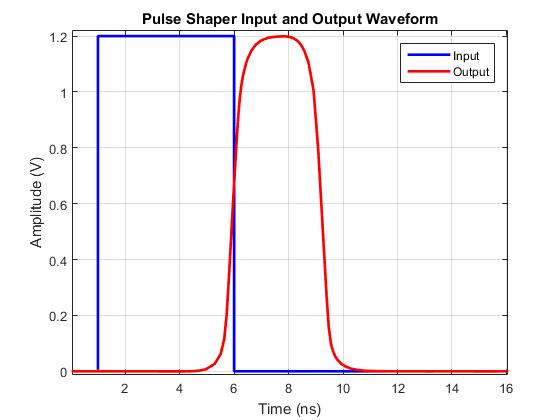} }%
    \label{ps_result}}
        \hspace{-0.9cm}
    \subfloat[\centering  ]{{\includegraphics[scale=0.26]{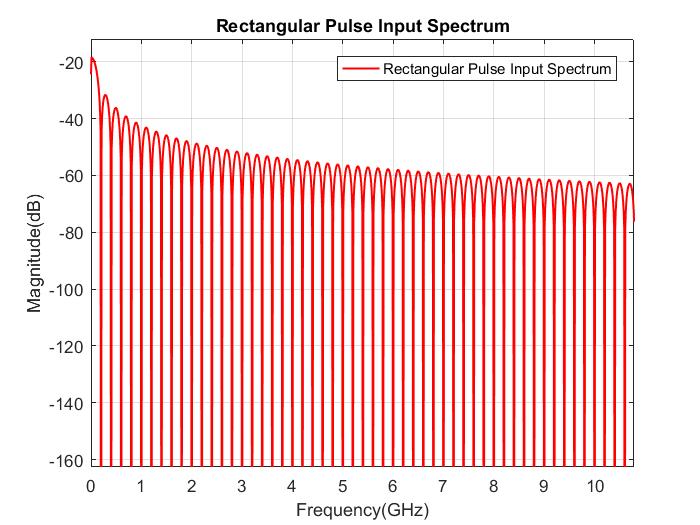} }%
    \label{ps_in}}
    \hspace{-0.9cm}
    \subfloat[\centering]
    {{\includegraphics[scale=0.26]{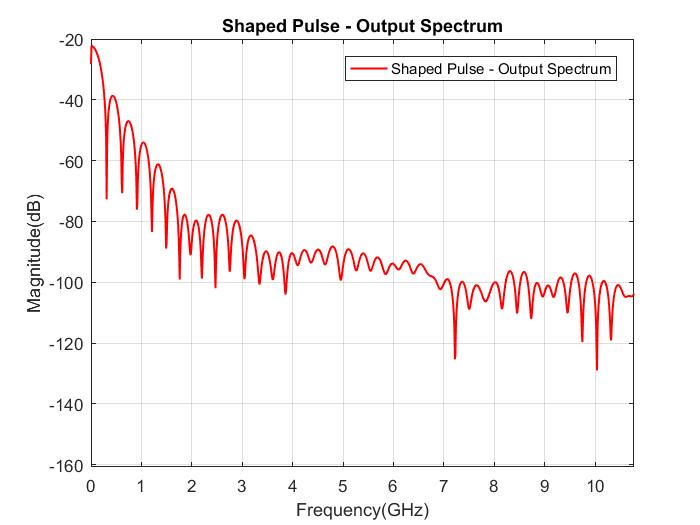} }%
    \label{ps_out}}
    \qquad
    \caption{  \protect\subref{ps_result} Pulse Shaping Circuit waveform for a pulse width of 5ns; \protect\subref{ps_in} Pulse Shaping Circuit Input Spectrum; \protect\subref{ps_out} Pulse Shaping Circuit Output Spectrum.}%
    \label{kernels}%
\end{figure*}


\subsection{Digitally Controlled Oscillator(DCO)}

The Digitally Controlled Oscillator (DCO)\cite{main_paper} generates sinusoidal signals in the 2–6 GHz range, with discrete frequencies of 2, 4, or 6 GHz set by the digital input to the capacitor filter bank. Coarse tuning is handled by the capacitor bank, and fine tuning by the varactor $C_{var}$. Fig.\ref{dco} shows the DCO circuit.

At these frequencies, parasitic capacitance significantly affects performance, as small changes cause large frequency shifts. The parasitic capacitance from MOS transistors M1, M2, M3, and M4 across the inductor is illustrated in Fig.\ref{vco_par}.

\begin{figure*}[!]%
    \centering
    \subfloat[\centering  ]{{\includegraphics[scale=0.263]{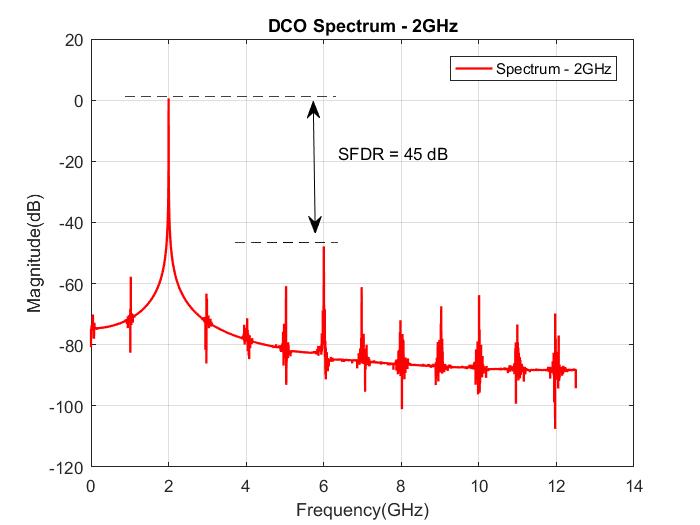} }%
    \label{dco_result1}}
            \hspace{-1.cm}
    \subfloat[\centering  ]{{\includegraphics[scale=0.263]{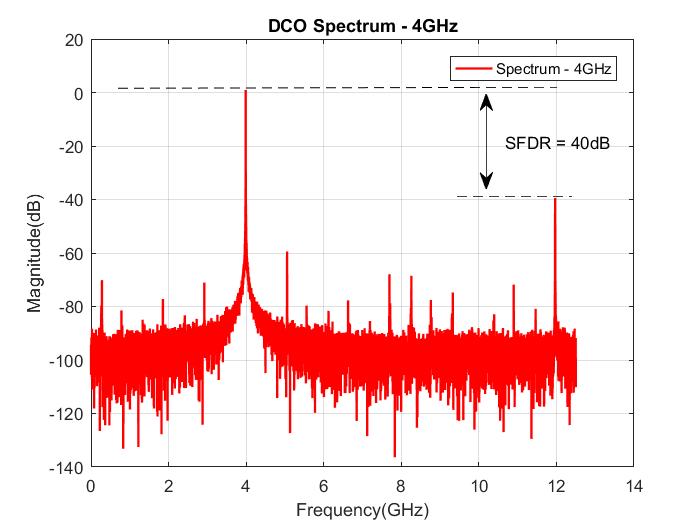} }%
    \label{dco_result2}}
            \hspace{-1.cm}
    \subfloat[\centering]
    {{\includegraphics[scale=0.263]{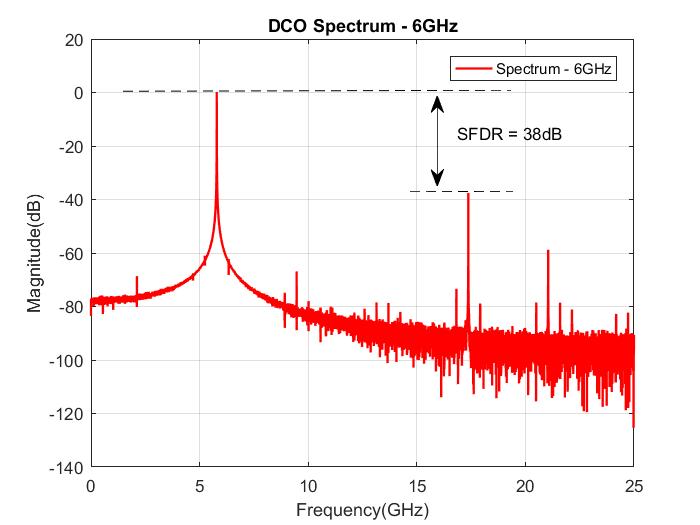} }%
    \label{dco_result3}}
    \qquad
    \caption{  \protect\subref{dco_result1} DCO Output Spectrum - 2 GHz; \protect\subref{dco_result2} DCO Output Spectrum - 6 GHz; \protect\subref{dco_result3} Digitally Controlled Oscillator Results.}%
    \label{kernels}%
\end{figure*}



The equivalent total parasitic capacitance $C_{par}$ across the inductor L can be expressed as,

\begin{dmath}
    C_{par} = \frac{(2C_{g_{d_4}} + C_{g_{s_4}})(2C_{g_{d_3}} + C_{g_{s_3}})}{2C_{g_{d_4}} + 2C_{g_{d_3}} + C_{g_{s_4}} + C_{g_{s_3}}} 
    + \frac{(2C_{g_{d_2}} + C_{g_{s_2}})(2C_{g_{d_1}} + C_{g_{s_1}})}{2C_{g_{d_2}} + 2C_{g_{d_1}} + C_{g_{s_2}} + C_{g_{s_1}}}
\end{dmath}

In general, the oscillation frequency of the oscillator can be written as,
\begin{equation}
    \omega_{o} = \frac{1}{\sqrt{L(C+C_{par})}}
\end{equation}

The parasitic capacitance of switches significantly affects the LC resonator core's total capacitance, with $C_{gd}$ determined by transistor size. A trade-off exists: larger switches reduce turn-on resistance and enable higher voltage swings but increase parasitic capacitance, while smaller switches minimize parasitics but limit voltage swing. Additionally, parasitic capacitance and channel resistance ($r_{ds}$) degrade the nMOS switch's quality

\subsubsection{Design Considerations for DCO} \hfill\\
\par For on-chip implementation, the inductor's finite quality factor (Q) introduces a lossy parallel resistor $R_p = \omega L Q$. This loss is compensated by a cross-coupled nMOS-pMOS pair, which provides a negative resistance of -$\frac{1}{g_m}$. To ensure compensation, $g_m$ should exceed $\frac{1}{R_p}$. Transistor sizing and the DC current ($I_{dc}$) are chosen based on this condition. This architecture offers twice the amplitude compared to nMOS-only or pMOS-only cross-coupled structures. The amplitude of the resulting sine wave is given by,

\begin{equation}
    A_{sine} = \frac{4.I_{dc}.R_{p}}{\pi}
\end{equation}

\subsection{Up-Conversion Mixer}
The Up-Conversion Mixer uses a double-balanced Gilbert Cell Mixer, which utilizes its symmetrical topology to cancel unwanted LO components from the RF. The well-matched adjacent components in the integrator circuit ensure high suppression of undesired signal components.


\section{Receiver}

The receiver front-end features a UWB LNA in a cascode common-source configuration with inductive degeneration, optimizing noise figure, gain, and power consumption. The proposed design is illustrated in Fig. \ref{LNA_C}\cite{lna_matching}.
\FloatBarrier

\par The circuit utilizes transistors M1 and M2 in a cascode common-source LNA structure. A three-section doubly terminated Chebyshev filter resonates the input impedance across the 2–6 GHz band

\begin{figure*}[!t]
  \centering
  \subfloat[]{\includegraphics[width=0.355\linewidth]{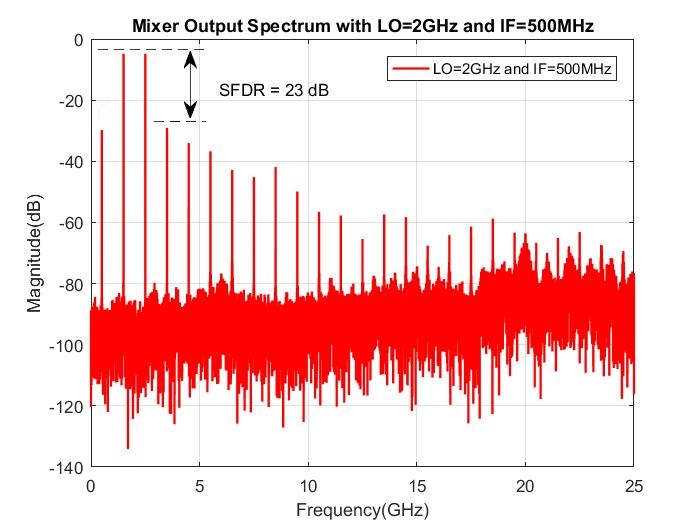}
		\label{mixer_result1}}
\hspace{-0.9cm}
 \subfloat[]{\includegraphics[width=0.355\linewidth]{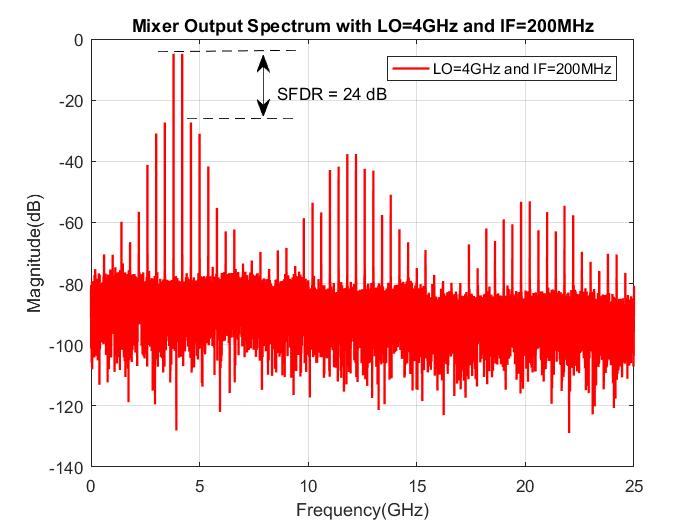}
		\label{mixer_result2}}
    \hspace{-0.9cm}
	\subfloat[]{\includegraphics[width=0.355\linewidth]{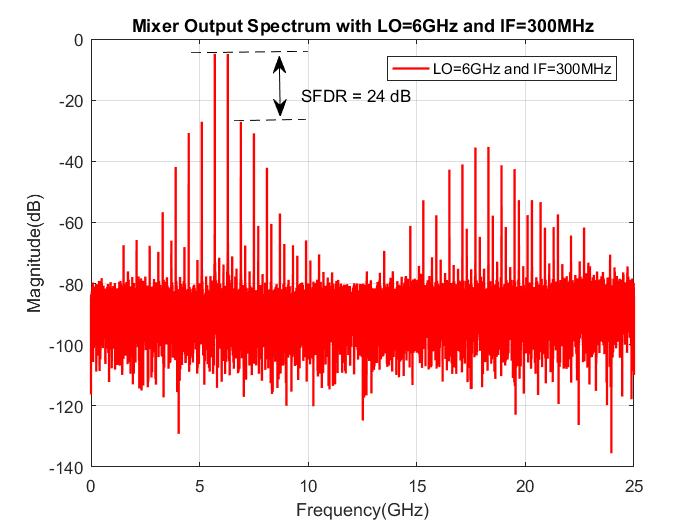}
		\label{mixer_result3}}
        
  	\subfloat[]{\includegraphics[width=0.355\linewidth]{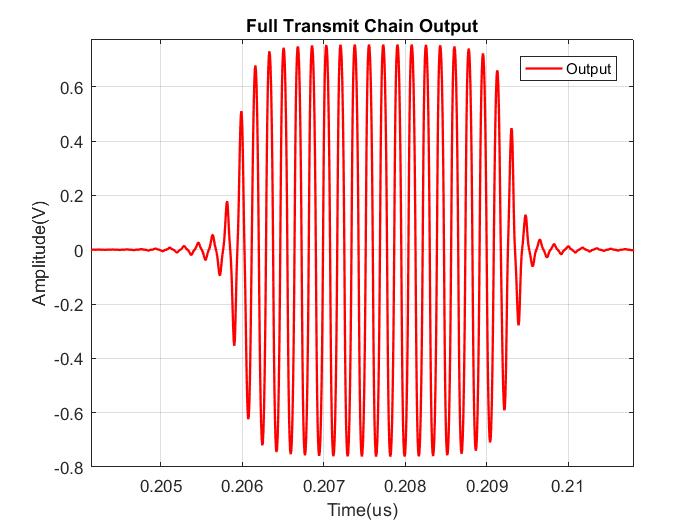}
		\label{tx_result1}}  
            \hspace{-0.9cm}
	\subfloat[]{\includegraphics[width=0.355\linewidth]{{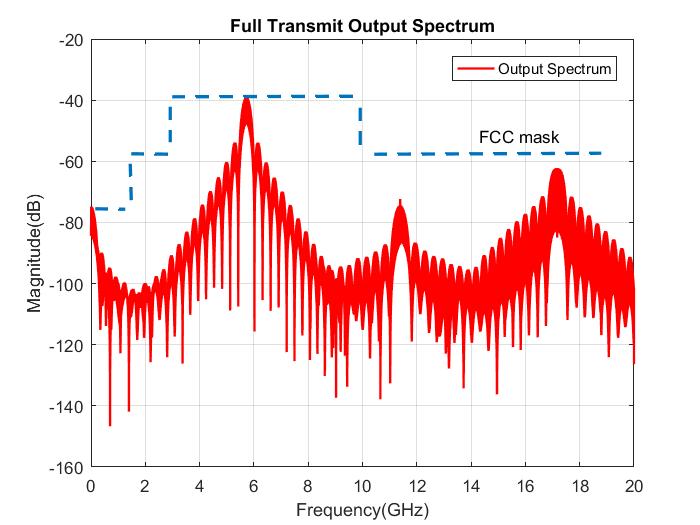}}
		\label{tx_result2}}
     \hspace{-0.97cm}
  	\subfloat[]{\includegraphics[width=0.356\linewidth]{{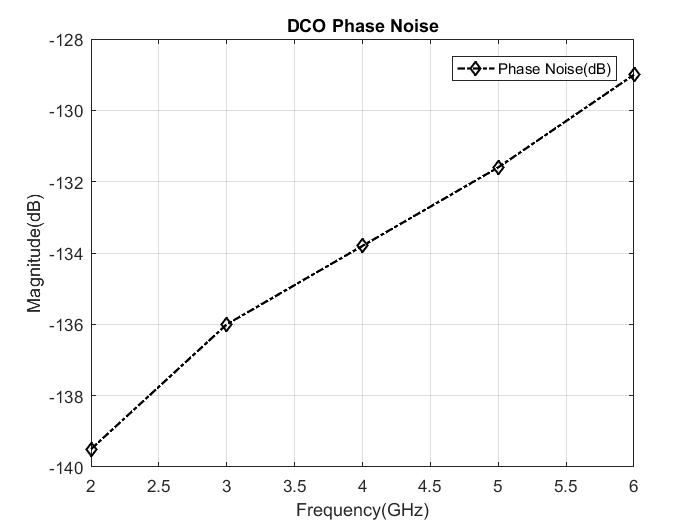}}
		\label{pn_dco}}
  \caption{\protect\subref{mixer_result1}~Mixer Output Spectrum - IF = 500MHz and LO = 2GHz. \protect\subref{mixer_result2}~Mixer Output Spectrum - IF = 200MHz and LO = 4GHz; 
 \protect\subref{mixer_result3}~Mixer Output Spectrum - IF = 300MHz and LO = 6GHz; \protect\subref{tx_result1}~Time domain waveform for Full Transmit Chain; \protect\subref{tx_result2}~Full Transmit Chain Output Spectrum; \protect\subref{pn_dco}Phase noise of the Digitally Controlled Oscillator}
  \label{taucellf}
\end{figure*}

\subsubsection{Input Matching} \hfill\\ 

\par The input impedance of the MOS transistor with inductive
source degeneration can be written as
\begin{equation}
  Z_{in}(s) = \frac{1}{s(C_{gs} + C_p)} + s(L_s + L_g) + \omega_T L_s
\end{equation}
where
\begin{equation}
  \omega_T L_s = \frac{g_m}{s(C_{gs} + C_p)}
\end{equation}

The inductive source degeneration network provides narrow-band matching, requiring additional circuitry for wide-band matching. The proposed design uses a Chebyshev filter at the input (Fig. \ref{LNA_i}), matching the real part of $Z_{\text{in}}$ to the source resistance $R_s$ (50 $\Omega$), equivalent to antenna impedance, over the 2–6 GHz range. Reactive elements in the filter determine bandwidth and in-band ripple, with the input reflection coefficient $\Gamma$ linked to ripple $\rho_p$\cite{lna_paper}.
\begin{equation}
  {\Gamma}^2 = 1 - \frac{1}{\rho_p} 
\end{equation}
 The gate-drain capacitance of M1 ($C_{gd}$) complicates analysis, introducing two series and one parallel resonance in the input matching network. One series resonance arises from $L_{g}$ and the combined capacitance of $C_{gd}$ with $L_{s}$ and $C_{p} + C_{gs}$. Parallel resonance occurs between $L_{s}$ and $C_{gd}$, while another series resonance involves $L_{g}$ and the parallel combination of $L_{s}$ and $C_{gd}$ at higher frequencies\cite{lna_matching}. The center frequency and bandwidth are tuned by adjusting $L_{1}$, $C_{1}$, $L_{2}$, and $C_{2}$.

\subsubsection{Gain Matching} \hfill\\
\par The input transistor's transconductance ($g_m$) primarily determines the amplifier's gain. In the LNA, transistors M1 and M2 form a cascode structure, enhancing small-signal gain, reducing the Miller effect, improving reverse isolation, and increasing output impedance by $g_{m2}r_{o2}$. The effective transconductance ($G_m$) is dominated by M1. The LNA voltage gain is given by the equation\cite{lna_matching}:
\begin{equation}
  \frac{v_{out}}{v_{in}} = -\frac{g_mW(s)}{s(C_{gs} + C_p)R_s} \frac{R_d\left(1+\frac{sL_d}{R_d}\right)}{1 + sR_dC_{out} + s^2L_dC_{out}}
\end{equation}
$W(s)$ represents the Chebyshev filter transfer function, while $L_d$, $R_d$, and $C_{out}$ are the load inductance, resistance, and total capacitance at M2's drain, respectively. The capacitance forms a spurious resonance with $L_d$, which must remain out of band. Adjusting $L_d$ allows gain variation.

\section{Experimental Results}
The radar transceiver circuit was implemented in 65nm CMOS technology. Transmitter components, including the DCO, mixer, and pulse-shaping circuit, were individually tested for functionality, with output waveforms and frequency spectra measured across 2–6 GHz. Receiver tests included LNA input matching, power gain, and noise figure measurements.

\subsection{Pulse Shaping Circuit}

Fig. \ref{ps_result} shows a 5 ns rectangular pulse input to the pulse shaping circuit and the resulting Gaussian-shaped pulse. Performance analysis in the frequency domain reveals that the FCC mask requires out-of-band emissions below -60 dB at 3 GHz. The pulse spectrum must drop below -60 dB at 1 GHz. As shown in Fig. \ref{ps_in} and Fig. \ref{ps_out}, the rectangular pulse achieves -44 dB, while the Gaussian pulse achieves -56 dB, making it more compliant with FCC regulations.


\subsection{Digitally Controlled Oscillator (DCO)}
Figs. \ref{dco_result1}–\ref{dco_result3} show the DCO frequency spectrum at 2, 4, and 6 GHz, with an SFDR exceeding 38 dB across the band. Linearity decreases with frequency (from 45 dB to 38 dB), with the third harmonic as the dominant spur, while the second harmonic is suppressed due to the differential DCO output.

Fig. \ref{pn_dco} plots DCO phase noise at a 1 MHz offset, showing an increase with frequency. This is attributed to parasitic MOS capacitance, $C_{par}$, varying non-linearly with signal amplitude, causing jitter and increased phase noise at higher frequencies. The phase noise arises from $\omega$ fluctuations in $\omega = \frac{1}{\sqrt{L(C+C_{par})}}$ due to $C_{par}$ becoming significant at higher frequencies. Additional effects influence phase noise, but this model aligns with the simulation results.

Key DCO performance parameters, including worst-case SFDR and phase noise at a 1 MHz offset, are summarized in Table \ref{t1}.

\begin{table}[h]
\caption{DCO performance parameters}
\label{t1}
\begin{center}
    
 \begin{tabular}{|c| c| c |} 
 
 \hline
 Sl.No & Parameter & Value  \\ [0.5ex] 
 \hline
 1 & SFDR & 38 dB\\ 
 \hline
 
 2 & Phase Noise$^{*}$ & -128 dBc/Hz\\ 
 \hline
 
 3 & Amplitude & 1.1 V\\
 \hline
 
 4 & Tuning Range & 2 - 6 GHz\\
\hline

 5 & Power & 6 mW\\ 
 \hline

 \end{tabular}
 
\end{center}
\end{table}


\begin{figure*}[!]
 \centering
 \subfloat[]{\includegraphics[width=0.355\linewidth]{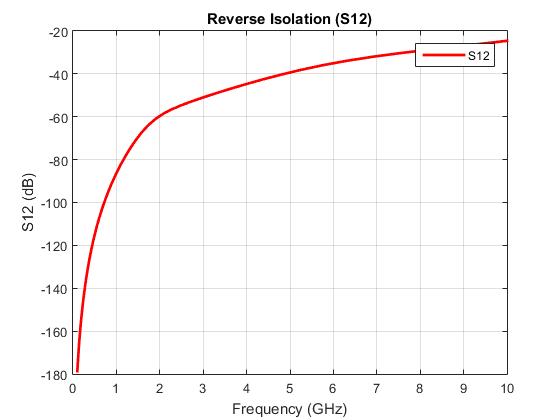}
		\label{iso}}  
    \hspace{-0.9cm}
 \subfloat[]{{\includegraphics[width=0.355\linewidth]{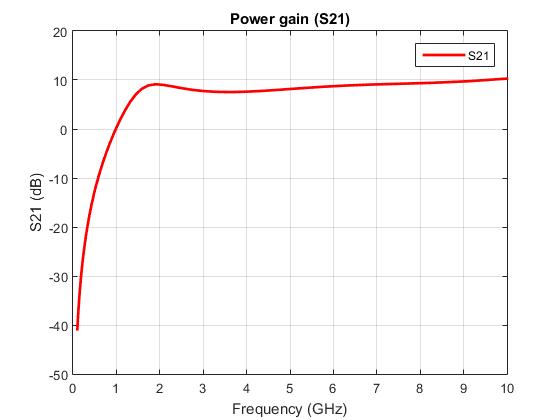}}
		\label{gain}}  
    \hspace{-0.9cm}
   \subfloat[]{{\includegraphics[width=0.355\linewidth]{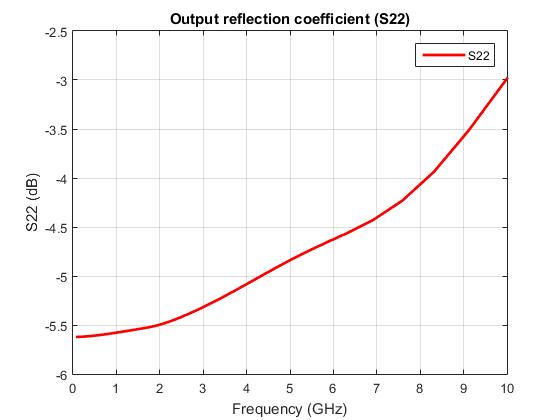}}
		\label{Or}}
        
          \subfloat[]{{\includegraphics[width=0.52\linewidth]{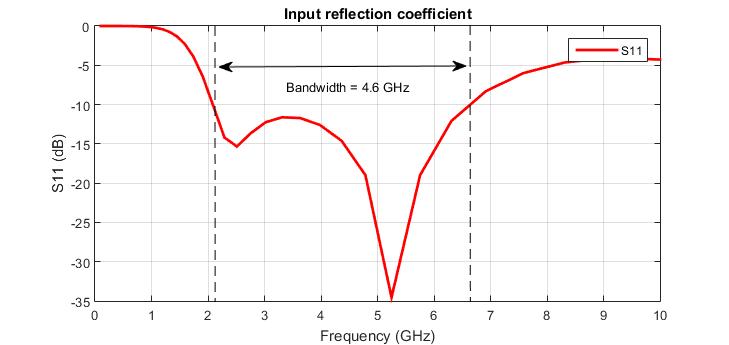}}
		\label{ir}}  
\subfloat[]{{\includegraphics[width=0.52\linewidth]{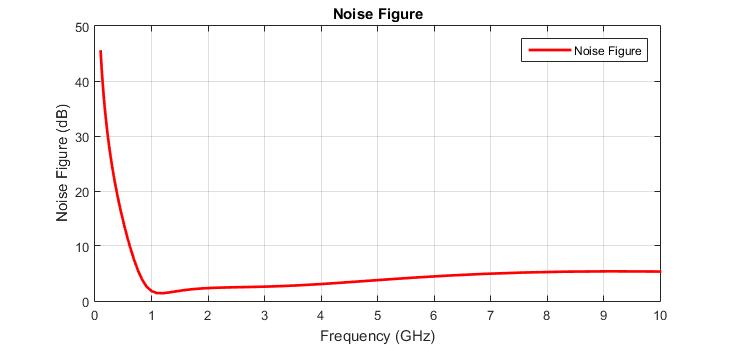}}
		\label{nf}}
  \caption{ \protect\subref{ir}Simulated input reflection coefficient
  \protect\subref{iso}Simulated reverse isolation; \protect\subref{gain}Simulated power gain \protect\subref{Or}-Simulated output reflection coefficient; \protect\subref{nf} LNA noise figure}
  \label{noisearray}
\end{figure*}


\subsection{Mixer}
Figs. \ref{mixer_result1}–\ref{mixer_result3} show the mixer output spectrum. Ideally, the output would contain only $\omega_{LO} + \omega_{IF}$ and $\omega_{LO} - \omega_{IF}$ components. However, intermodulation products from odd harmonics of the IF and LO frequencies appear due to the inherent non-linearity of the nMOS device. Key mixer performance parameters are summarized in Table \ref{t2}.

\begin{table}[h]
\caption{Mixer performance parameters}
\label{t2}
\begin{center}
    
 \begin{tabular}{|c| c| c |} 
 
 \hline
 Sl.No & Parameter & Value  \\ [0.5ex] 
 \hline
 1 & SFDR & 23 dB\\ 
 \hline
 
 2 & Noise Figure & 11.2 dB\\ 
 \hline
 
 3 & Conversion Gain & 1.2 dB\\
 \hline
 
 4 & Power & 12 mW\\
\hline

 \end{tabular}
 
\end{center}
\end{table}


\subsection{Full Transmit Chain}

Fig. \ref{tx_result1} shows a 5 ns pulse with a 6 GHz center frequency, displaying a smooth amplitude build-up due to the Gaussian pulse shaping. Fig. \ref{tx_result2} presents the corresponding spectrum, which remains below the FCC mask in the 2-6 GHz band and other bands, ensuring compliance with FCC outdoor transmission regulations.

   

\subsection{Low Noise Amplifier (LNA)}

The LNA simulation results include S-parameters and noise figure (NF). The S-parameters—$S_{11}$ (input reflection), $S_{12}$ (reverse isolation), $S_{21}$ (power gain), and $S_{22}$ (output reflection)—are analyzed. Fig. \ref{ir} shows the input matching with $S_{11}$ below -10 dB across the 2.1-6.7 GHz UWB band, achieving excellent matching, with the best at 5.25 GHz (-35 dB). Reverse isolation $S_{12}$ is better than -30 dB, minimizing LO leakage. Power gain $S_{21}$ ranges from 8 dB to 10 dB (Fig. \ref{gain}), and output matching is shown in Fig. \ref{Or}


The simulated noise figure (NF) of the LNA, shown in Fig. \ref{nf}, ranges from a minimum of 2.5 dB at 2.1 GHz to a maximum of 5 dB at 6.7 GHz, with a power consumption of 11 mW. The component values in the matching network (Fig. \ref{LNA_C}) are listed in Table \ref{t3}.


\begin{table}[h]
\caption{LNA matching network circuit components}
\label{t3}
\begin{center}
    
 \begin{tabular}{|c| c| c |} 

 \hline
 Sl.No & Circuit component & Value  \\ [0.5ex] 
 \hline
 1 & $L_1$ & 1.375 nH\\ 
 \hline
 
 2 & $C_1$ & 1.08 pF\\ 
 \hline
 
 3 & $L_2$ & 3 nH\\
 \hline
 
 4 & $C_2$ & 0.1 pF\\
\hline

 5 & $L_g$ & 0.2 nH\\ 
 \hline
 
 6 & $C_p$ & 1.38 pF\\ 
 \hline
 
 7 & $L_s$ & 1 nH\\ 
 \hline

 8 & $L_d$ & 1.75 nH\\ 
 \hline

 9 & $R_d$ & 40 {$\Omega$}\\ 
 \hline

 \end{tabular}
 \end{center}
 \end{table}
\section{Conclusion}
The design of a CMOS transceiver front-end implemented in 65nm technology is presented. The system operates in the frequency range of 2-6GHz with ultra-wideband frequency tuning capability. The FCC spectrum mask specifies the maximum power that can be transmitted in different frequency bands and the transmitted signal adheres to the FCC spectrum mask. At the receiver side, the LNA with Chebyshev matching shows advantages in overall performance (NF, power gain, large bandwidth), compared to the distributed, conventional shunt-feedback or filter-based amplifiers that make up other wide-band topologies. The application for this system can be found mainly in the areas including through wall detection and bio-medical applications. The frequency diversity capability, along with the low-cost and the high integration of the proposed UWB radar transceiver, makes it a very attractive and cost-effective SoC solution for low-power radar sensing applications.

  
  
%

\ifCLASSOPTIONcompsoc
  \section*{Acknowledgments}
\else
\fi


\ifCLASSOPTIONcaptionsoff
  \newpage
\fi



%

%

\begin{IEEEbiography}{Michael Shell}
Biography text here.
\end{IEEEbiography}

\begin{IEEEbiographynophoto}{John Doe}
Biography text here.
\end{IEEEbiographynophoto}


\begin{IEEEbiographynophoto}{Jane Doe}
Biography text here.
\end{IEEEbiographynophoto}




\end{document}